\documentclass[11pt,a4paper]{article}
\usepackage[utf8]{inputenc}
\usepackage[T1]{fontenc}
\usepackage[british]{babel}
\usepackage{amsmath, amsfonts, mathtools, amssymb, graphicx, float, lipsum, xcolor, hyperref, authblk, multicol, framed, amsthm}
\usepackage{algorithm, algpseudocode}
\hypersetup{hidelinks}
\usepackage[left=2.00cm, right=2.00cm, top=2.00cm, bottom=2.00cm]{geometry}
\graphicspath{{./figures/}}

\usepackage{array}
\usepackage{booktabs}
\usepackage{cite}
\usepackage{tikz}
\usepackage{float}
\usepackage{tabularx}
\usepackage{adjustbox}
\usepackage{longtable}
\usepackage{rotating}

\newtheorem{example}{Example}[section]
\newtheorem{definition}{Definition}[section]
\DeclareMathOperator{\AVP}{AVP}
\newcommand{\bigO}{\mathcal{O}}

\title{BATTLE for Bitcoin: Capital-Efficient Optimistic Bridges with Large Committees}
\author[1,2]{Sergio Demian Lerner\thanks{\href{mailto:sergio@fairgate.io}{sergio@fairgate.io}}\hspace{3pt}} % \author[]{}

\author[1]{Ariel Futoransky\thanks{\href{mailto:futo@fairgate.io}{futo@fairgate.io}}\hspace{3pt}} % \author[]{}
\affil[1]{Fairgate Labs} % \affil[]{}
\affil[2]{Rootstock Labs}
\date{}

\usepackage{xspace}

\newcommand{\txname}[1]{\texttt{#1}}

\newcommand{\AMIC}[1][]{\ensuremath{A_{\mathsf{MIC}}^{\mathsf{#1}}}\xspace}

\newcommand{\ADR}[1][]{\ensuremath{A_{\mathsf{DR}}^{\mathsf{#1}}}\xspace}

\newcommand{\AOSB}[1][]{\ensuremath{A_{\mathsf{OSB}}^{\mathsf{#1}}}\xspace}

\newcommand{\APSB}[1][]{\ensuremath{A_{\mathsf{PSB}}^{\mathsf{#1}}}\xspace}

\newcommand{\DROCA}{DROCA }

\begin{document}
%\tableofcontents
\maketitle

\begin{abstract}
We present \emph{BATTLE for Bitcoin}, a DoS-resilient dispute layer that secures
optimistic bridges between Bitcoin and rollups or sidechains. Our design adapts the
BATTLE tournament protocol to Bitcoin’s UTXO model using BitVM-style \textsf{FLEX}
components and garbled circuits with on-demand L1 security bonds. Disputes are resolved
in logarithmic rounds while recycling rewards, keeping the honest asserter’s minimum
initial capital constant even under many permissionless challengers. The construction is
fully contestable—challengers can supply higher-work counter-proofs—and relies only on
standard timelocks and pre-signed transaction DAGs, without new opcodes.

For $N$ operators, the protocol requires $\mathcal{O}(N^2)$ pre-signed transactions,
signatures, and message exchanges, yet remains practical at $N\!\gtrsim\!10^3$, enabling high decentralization. 

\end{abstract}

\section{Introduction}

The BATTLE~\cite{battle} protocol provides a mechanism to resolve multiparty disputes on a Turing-complete blockchain. In this work we show how to implement the BATTLE protocol on Bitcoin\cite{bitcoin_whitepaper}, and how it can be used to secure an optimistic bridge. Existing optimistic bridge designs that protect both provers and challengers from resource
exhaustion typically rely on outcome-dependent, dynamic dispute scheduling—feasible on
stateful VMs but not directly on Bitcoin’s stateless UTXO model and constrained scripting.
We replicate the needed dynamics on Bitcoin using pre-signed transaction DAGs, timelocks,
and on-demand bonds, incurring only minor trade-offs.

Another difficulty in the UTXO model is transaction-level concurrency. If assertions can
be posted asynchronously while sharing the same dispute window, no static schedule can
guarantee challenger protection against resource-exhaustion. To address this, we introduce
the \emph{Tournament Chain} (TC), which serializes tournament openings (via rate-limited
links) to prevent concurrent asynchronous assertions and restore challenger protection.

While protecting both challengers and asserters is desirable, in the bridge setting, it can suffice to harden
the protocol primarily for asserters. A deployment may, for example, operate with a closed
set of about $20$ asserters alongside a permissionless population of roughly $200$
watchtowers and remain meaningfully trust-minimized: asserters provide liveness and
censorship-resistance by submitting peg-out claims, whereas watchtowers enforce safety by
disputing invalid claims. In practice, asserters are typically liquidity providers that
front funds to users and later seek reimbursement—an activity concentrated among a few
specialized firms—whereas watchtowers can be individuals, nonprofits, or DAOs associated
with applications. Thus, the asserter set is naturally small, while the challenger set can
be large and open.

\subsection{Contributions}

BATTLE for Bitcoin is a concurrent-assertion dispute resolution protocol (DROCA~\cite{battle}) that achieves logarithmic-time dispute resolution for Bitcoin, while keeping the honest asserter’s on-demand security bonds constant. The protocol proceeds in rounds: in round $r$ it opens $k_r$ disputes, with a non-decreasing schedule $k_{r+1}\ge k_r$ (typically $k_{r+1}\approx 2k_r$). Dispute rewards are paid before the next round and are earmarked to fund the security bonds and fees of round $r+1$, enabling exponential growth in per-round concurrency and thus $\bigO(\log C)$ rounds to resolve $C$ challengers.

A large number of malicious provers can overwhelm the capacity of an honest challenger to challenge them all. Recent research on this topic includes the PRT~\cite{nehab2022prt}, BoLD~\cite{alvarez2024bold}, Dave~\cite{dave_fraudproof} and Optimism~\cite{optimism_fraudproof} protocols. These protocols run on blockchains with Turing-complete stateful smart contracts such as Ethereum\cite{wood2014ethereum} or Rootstock~\cite{rsklerner22}. However, the main focus of our work is bridging the Bitcoin blockchain\cite{bitcoin_whitepaper}, which is natively stateless and has a restricted scripting language. Existing solutions that protect both provers and challengers from resource exhaustion require a dynamic schedule of disputes based on past outcomes, and while this is possible on Ethereum, it is not possible on Bitcoin. 

The BATTLE for Bitcoin protocol does not exactly conform to the \DROCA abstraction. A minor deviation is that an asserter defending a claim may submit a different assertion in distinct
two–party disputes. This flexibility is benign: if the original assertion is correct,
replacing it with an incorrect one only weakens the asserter’s position; if it is
incorrect, switching to another incorrect assertion offers no advantage. Thus, allowing
per–dispute substitutions does not improve the asserter’s chance of success.

\subsection{BATTLE}

BATTLE is a dispute-resolution protocol for the DROCA setting that converts adversarial
\emph{concurrency} into bounded \emph{time} while keeping honest asserter capital
requirements constant. It organizes all disputes for a given sequence number into a
two–phase tournament executed in rounds:

\emph{Phase~1 (asserter bracket).} All sibling assertions first compete in a bracket that
eliminates conflicting asserters until a single surviving assertion remains. Each round
completes within a fixed time bound; winners receive a dispute reward taken from the
loser’s on–demand bond.

\emph{Phase~2 (challenger escalation).} The surviving asserter then faces all registered
challengers. Rounds are scheduled so that the number of simultaneous disputes per round
is non–decreasing (e.g., maintain–or–double or gradual schedules). Rewards won in one
round are \emph{recycled} to fund bonds and fees in subsequent rounds (progressive
buy-in), yielding \(O(\log C)\) worst-case rounds against \(C\) challengers with
\emph{constant} minimum initial capital for the honest asserter.

Economically, BATTLE distinguishes: (i) a per-dispute \emph{cost of dispute} (fees and
off-chain work), (ii) \emph{on-demand security bonds} posted for each two-party dispute,
(iii) optional \emph{persistent bonds} for slashing, and (iv) \emph{dispute rewards}
(paid from losing bonds and immediately reusable next round). The protocol assumes
bounded-time, black-box two-party verification games and enforces per-move deadlines so
that disputes do not overlap indefinitely and also assumes external limits on assertion creation. We instantiate such black-box two-party disputes on Bitcoin using FLEX~\cite{FLEX} components and provide limits the assertion creation rate with a Tournament Chain.

The BATTLE protocol specifies an \emph{Assertion Verification Predicate} (AVP) that,
given an assertion and fixed parameters~$\theta$, returns $1$ iff the assertion is
correct under~$\theta$, and $0$ otherwise, within a bounded time
$T_{\mathrm{avp}}$ (a protocol constant). In this work, use two equivalent realizations of the AVP: (i) a garbled-circuit~\cite{yao_garbled_circuits} AVP used on chain to conditionally
reveal secrets and settle disputes, and (ii) a programmatic AVP used off chain by
participants to rapidly screen assertions and decide whether to challenge. Both
descriptions must be semantically identical. Asserters can evaluate the $\AVP$ on their own candidate assertions off-chain prior to communication, so the predicate’s outcome is known to them before submission.

Some inputs to the predicate (e.g., the asserter identity) are fixed by the contract with restrictions established using Bitcoin script constraints. These restrictions could have been established by embedding the fixed inputs directly in the garbled circuit wires, however, this prevents reusing the same circuit template as-is. 

The BATTLE protocol defines two roles: \emph{asserters}, who are authorized to submit assertions and defend them in disputes, and \emph{watchtowers}, who are authorized to register and prosecute challenges. Parties may hold both roles. Parties are also known as \emph{operators}. 

BATTLE protocol splits time into discrete epochs (or rounds). We further subdivide each epoch  into five timelock periods, so that a FLEX dispute can be fully resolved in one epoch.

\subsection{BATTLE for Bitcoin Bridges}

BATTLE protocol was designed to serve as the dispute layer for optimistic Bitcoin bridges~\cite{paper:bitvm2bridge,union_bridge,clementine_bridge,bal2025clementine,strata_alpenlabs,linus2024bitvm2bridge}. In this paper, we focus on that deployment setting. Existing deployments are instantiated atop the BitVM family of protocols~\cite{paper:bitvm,aumayr2024bitvm,bitvmx_paper,paper:bitsnark}. In these designs, security bonds are not posted at \emph{claim time}; instead, they must be provisioned at \emph{setup time} within pre-signed transaction graphs, which increases capital lockup and operating cost. The absence of on-demand bonding reduces capital efficiency and raises the minimum initial capital required of operators.

BitVM schemes instantiated with garbled circuits~\cite{glock25}, together with the FLEX protocol~\cite{FLEX}, enable \emph{on-demand} security bonds for peg-out assertions. This capability was not available in earlier Bitcoin-based designs, where bonding had to be pre-provisioned at setup time. Consequently, prior to FLEX, resource-efficient tournament scheduling on Bitcoin was impractical.

\subsection{Connection to Multiparty Lotteries on Bitcoin}

The Bitcoin-specific Phase~1 of BATTLE closely parallels prior \emph{Multiparty Lottery}
(ML) protocols on Bitcoin~\cite{miller17}. In ML, \(n\) participants escrow stakes into a
common pot, and a uniformly (pseudo)randomly selected winner receives the entire pot. ML protocols have been studied in blockchain settings—and specifically on Bitcoin—focusing on fair randomness generation, robustness against aborts, and efficient on-chain settlement~\cite{Bartoletti16,miller17}.
 
The most capital-efficient ML constructions on Bitcoin eliminate per-party security deposits but require an exponential (in the number of participants $n$) quantity of pre-signed transactions~\cite{miller17}.

Both \DROCA and ML are multi-party computation (MPC) settings, but \DROCA is strictly more expressive; in particular, an ML can be realized as a special case of \DROCA. One reduction is to let each participant post a sibling assertion and then resolve them by pairwise elimination using a two-party lottery. The two-party lottery follows a commit–reveal pattern: Alice and Bob first commit to random seeds $a,b$ (e.g., by posting $H(a)$ and $H(b)$), then reveal $a,b$, and a selector
\[
f(a,b) \;=\; \big(\mathrm{parity}(a)+\mathrm{parity}(b)\big)\bmod 2
\]
chooses the winner. A Bitcoin realization can use a FLEX-like component in which Alice and Bob prepare garbled circuits $C_A$ and $C_B$ that both take $(a,b)$ as input, verify the commitments, compute $f$, and release a winning secret: $C_A$ releases a secret redeemable by Bob if $f(a,b)=0$, while $C_B$ releases a secret redeemable by Alice if $f(a,b)=1$. Alternatively, the selector $f$ can be implemented directly in Bitcoin Script via a parity-of-length test on two input strings, yielding a script-level winner selection without garbled circuits~\cite{Marcin13}.

A BATTLE for Bitcoin can realize the ML primitive as a special case. In this reduction, per-party on-demand security bonds are \emph{constant} (independent of $N$), and the number of pre-signed transactions required for setup scales quadratically, $\bigO(N^2)$, rather than exponentially, $\bigO(2^N)$, as in prior Bitcoin ML constructions.

\section{Implementation}

Implementing BATTLE for Bitcoin requires \emph{on-demand} L1 security bonds that can be posted and released on a per-dispute basis. We realize this using FLEX-style BitVM garbled-circuit disputes and extend the FLEX component to support (i) per-move timelocks aligned with epoch scheduling, (ii) escrowed dispute-reward outputs that are immediately reusable in subsequent rounds, and (iii) race-resilient early-refund gates. These refinements preserve UTXO semantics and enable BATTLE’s capital-recycling tournament to execute natively on Bitcoin.

At first glance, one might attempt to post \AOSB\ on the side-system before requesting
reimbursement. This is ineffective: a dishonest asserter can simply avoid funding the
side-system bond and present a forged (or otherwise invalid) peg-out proof on Bitcoin,
seeking reimbursement without any enforceable collateral on L1. Because reimbursement
is executed on Bitcoin and cross-system enforcement is non-atomic, \AOSB\ must be
escrowed directly on the Bitcoin blockchain to make slashing and rewards binding.
By contrast, the persistent bond \APSB\ serves long-lived incentive purposes and can
be maintained on the side-system.

\subsection{The FLEX Component}

The FLEX component~\cite{FLEX} is a garbled circuit-based protocol designed to facilitate two-party disputes on Bitcoin without requiring permanent security bonds. FLEX enables conditional 'on-demand' security deposits that are only activated in the event of a dispute, reducing the financial overhead for operators and challengers. The main goal of FLEX is to improve the capital efficiency of BitVM-based bridges in a permissioned challenge setting but can also be used to improve the security of any other fraud proof-based protocol. In this work, we use multiple FLEX components either chained or concurrently to support multi-party tournaments.

\subsection{Assumptions for Bitcoin}

To protect BATTLE against Censorship, Ordering, Resource Exhaustion, and Delay Attacks, we need to make some assumptions.
We abstract the continuous nature of time and assume that time is divided into discrete epochs, all disputes can be fully resolved in one epoch (or 'round').

No chess clock is used: we assume that individual transaction timelock periods within the epoch are chosen so that censorship attacks are irrational: spamming the blockchain to exclude a transaction will cost more to the attacker than the value at stake to grab. Therefore, we assume that no delay attack will be sustained for a timelock period or more. We do consider the incidence of delay attacks of periods below a timelock period, and so we show that our protocol is free from race conditions that can emerge from short delays. 

We assume that the issuance of dispute transactions does not generate network or block congestion. Therefore, the disputes do not interfere with each other at the network level, nor compete for block space or block gas. 

We assume that honest parties can follow the protocol and act immediately when they need to. If a party needs to perform a local computation before issuing a transaction, then the global timelock period or chess clock is adjusted to account for the worst-case local computation time, added to the censorship resistance requirement. 

We assume that when there is no ongoing spamming attack, transactions are immediately included in the blockchain (the time it takes to the first confirmation is considered negligible compared to the timelock period). When implemented on UTXO-based blockchains, we assume that child transactions can be issued without confirmation of the parent, so if a party receives bitcoins by the end of one period, he is able to resend those bitcoins immediately at the beginning of the following period. These assumptions can be relaxed by adding additional timelock periods to each tournament round to ensure that reward transactions are confirmed before they need to be spent again.
We also assume that the UTXOs used are mature (they are not restricted by the maturity lock on recently mined bitcoins).

We assume that disputes do not carry extractable value to block producers. Nevertheless, our protocol is immune to transaction ordering attacks within the same block unless it happens in the boundary of a timelock period. 

To summarize, we assume parties do not delay their responses, local computation is accounted for in the time-lock period, and transactions are included in blocks immediately when they are broadcast. Unexpected transaction delays are entirely attributed to censorship attacks.

We care especially about resource exhaustion attacks against asserters and set our initial capital requirements so that the resources of honest parties cannot be exhausted. 

We assume nodes and participants employ standard
fee-bumping (RBF/CPFP) and re-broadcast policies to counter \emph{transaction replacement
cycling attacks} (TRCAs)~\cite{bitcoinoptech_replacement_cycling,riard2023,riard2025}.
We model TRCAs as imposing strictly positive, non-recoverable fee costs on the attacker
per cycle, while honest parties can rebroadcast and bump within the protocol’s timelocks.
Under this assumption, sustained censorship via TRCAs is economically irrational, and any
residual delay is accounted for within the cost-of-dispute parameter.

\subsection{Minimum Initial Capital and Concurrency}

In BATTLE, an honest asserter requires a \emph{constant} minimum initial capital (independent of the number of challengers $C$) to win all disputes,
and the worst-case number of rounds satisfies $R=\bigO(\log C)$. A comparable constant-capital
guarantee holds for challengers provided the system enforces a bound on the number of
\emph{simultaneously active tournaments}. Enforcing such a concurrency cap is
particularly challenging on Bitcoin, which lacks a reliable clock oracle, stateful
scheduling primitives, or robust block-time introspection. As an approximation, we
employ a \emph{rate-limited Tournament Chain (TC)} that admits a new tournament only
after a fixed relative timelock expires or upon agreement between all parties; while effective
at reducing concurrency, this mechanism is coarse and may penalize throughput. In the
absence of any cap, a challenger that aims to contest all active assertions must
provision capital that grows linearly with the number of simultaneous tournaments.

\subsection{Setup}

During setup, parties communicate over encrypted and authenticated point-to-point
channels. Each participant distributes its public keys (e.g., Schnorr for on-chain
spends and Winternitz for one-time revelations), along with the hashes of transaction
templates and other commitments needed to assemble the transaction DAG. Parties then
construct their local view of the DAG, sign the transactions for which their signatures
are required, and exchange the resulting signatures so that counter-parties can
complete their own views. A complete global view is not required: each party may
retain only the subgraph relevant to its role.

To ensure consistency, shared DAG components are accompanied by collision-resistant
digests (e.g., per-subgraph hashes or Merkle roots). Parties verify the received
artifacts, compute local digests, and compare them with peers to confirm that the
overlapping portions of their DAGs match. The protocol proceeds only if all
participants acknowledge successful completion of setup and convergence on the
committed digests.

\subsection{Phase 1}

To preclude simultaneous reimbursement attempts at the same sequence number, we
introduce a Bitcoin-compatible Phase~1. The construction pre-commits to a dispute
DAG that includes a potential two-party dispute for every pair of registered
asserters (a fixed bracket), but only a schedule-selected \emph{subset} of these
disputes is activated on chain. Activation follows a predefined timetable (via
timelocks), so at any epoch only the slated pairwise matches can proceed, while
the remaining edges stay dormant and impose no on-chain footprint. 

All potential asserters are by default registered in the tournament. A party that does not wish to participate passively waits for the first a timelock period to expire.
In the first round, all parties will odd index numbers defend, while all parties with even index numbers chalenge.

If a defender party confronts a nonparticipating challenger, the defender party simply waits for one timelock period and issues the \txname{BobNoChallenge} transaction. If a challenger confronts a nonparticipating defender in the first round, then the challenger waits a timelock period and issues a \txname{AsserterTimeout} transaction. 

If two paired asserters both abstain, no interaction between them is required: any honest participant may issue \txname{DisputeTimeout} to retire the dormant winner selection edge and remove both parties from that bracket slot. In other rounds past the first, if a participant 
is paired against another participant that drops off during the tournament, the match is decided by walkover in favor of
the participant who stays. 

The Phase~1 tournament requires $\bigO(N^2)$ pre-signed transactions, and the total number of signatures is likewise $\bigO(N^2)$.

Disputes are scheduled and activated by relative timelocks, and each match must conclude
before its timelock expires. Upon resolution (including timeout), the loser—asserter or
challenger—is eliminated and cannot advance to subsequent rounds.

\begin{figure}[h]
\centering
\includegraphics[width=0.6\textwidth]{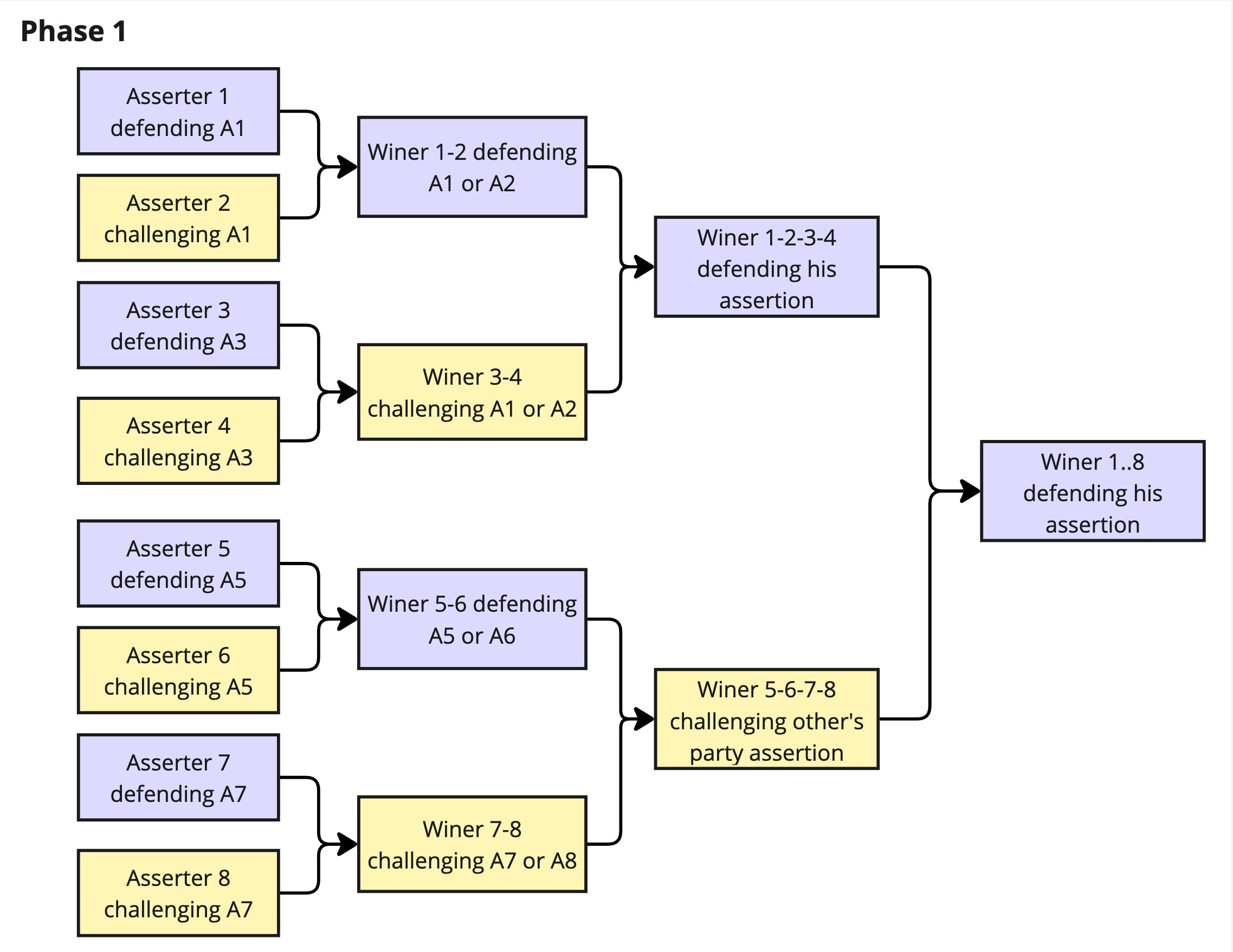}
\caption{A fixed phase 1 tournament schedule}
\label{fig:phase1-fixed}
\end{figure}

Figure ~\ref{fig:phase1-fixed} shows all potential matches from a phase 1 tournament.

\begin{figure}[h]
\centering
\includegraphics[width=0.6\textwidth]{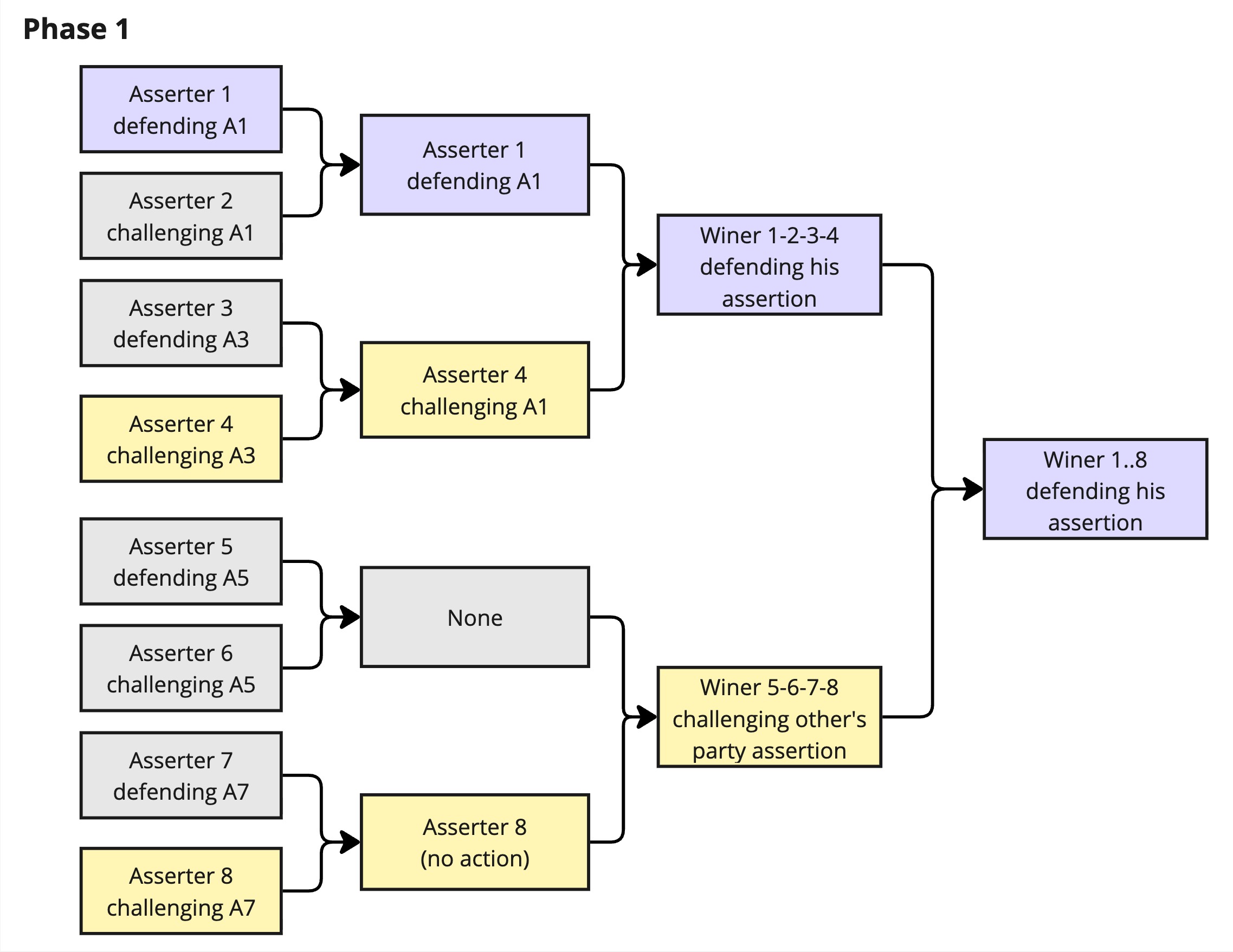}
\caption{The development of a phase 1 when only asserters 1,4 and 8 participate}
\label{fig:phase1-ongoing}
\end{figure}

Suppose only asserters $1$, $4$, and $8$ opt in (all others withdraw or time out). 
Figure~\ref{fig:phase1-ongoing} depicts the resulting Phase~1 progression under the predefined schedule.

If \AMIC\ is sufficient to fund $Q$ concurrent disputes per round, we can run $Q$
independent tournament brackets in parallel. Assume $Q=2^{k}$. Then each party plays
$Q$ matches per round (one per bracket), increasing the per-round elimination rate by
$2^{k}$ and reducing the wall-clock number of rounds from $R$ to $R-\log_{2} Q$.

\subsubsection{Registration}

All asserters holding sibling claims register by issuing a per-asserter registration
transaction \txname{EnableRound(1)}. Registration activates only the dispute scheduled for the next timelock epoch. 

\subsection{Enablement Chains}

To allow the participation of each party in the tournament rounds, we employ an \emph{enabler chain} per asserter: a time-ordered sequence of
links with outputs that authorize that asserter’s participation in successive rounds. Failure to challenge within the window allows any other participant to exclude $X$ by
broadcasting \txname{NoAssertion}$(X)$.
Each pairwise dispute is pre-signed so that the winner can \emph{cut} the loser’s next-link
enabler, thereby disabling the loser for future rounds without additional co-signatures.

Additionally, each winner gets to consume a winner selector output that has a relative timelock to the first transaction of the tournament. When the timelock expires, only one of the two participants in a dispute should have the capability to consume the winner selection, because tThe next \txname{EnableRound} transaction can only be issued by the participant that consumes the selector. There is one winner selector output for each match in the bracket.

Each link in the enabler chain (transaction \txname{EnableRound})$(i)$ is guarded by a relative timelock of six timelock
periods relative to the time the match should have started. This imposes a lower bound on optimistic completion: even if no additional
asserters register, Phase~1 cannot finalize in fewer than $6\cdot \lceil \log_2 N\rceil$ timelock periods. An equivalent
gating effect could be realized via an “early-win” mechanism, but doing so would introduce additional branches in the transaction DAG; for clarity, we
present the simpler enabler-chain construction.

\begin{figure}[h]
\centering
\includegraphics[width=0.6\textwidth]{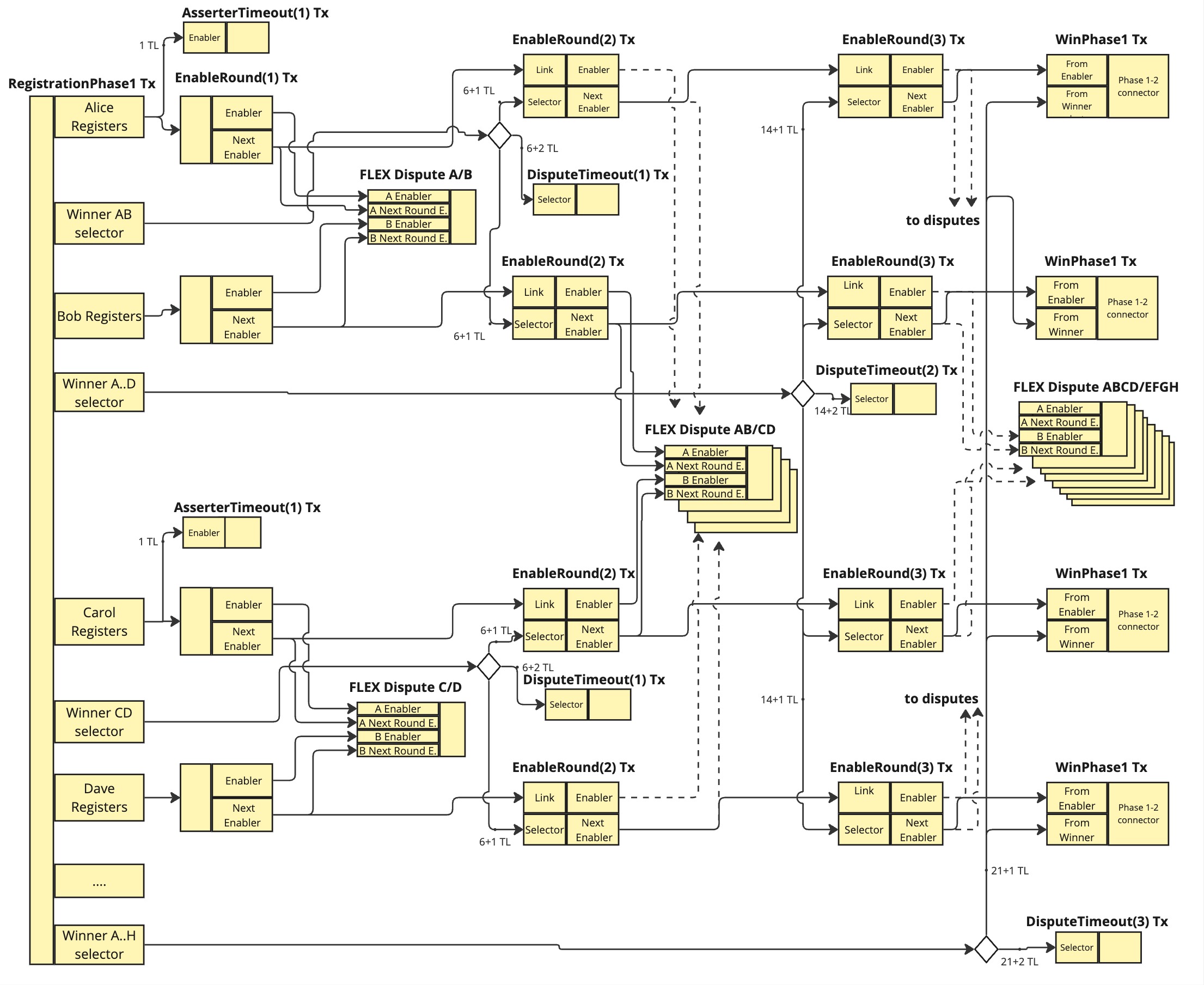}
\caption{A Phase 1 Registration, Enablement chains and FLEX disputes}
\label{fig:phase1-seq}
\end{figure}

Figure~\ref{fig:phase1-seq} depicts the Phase~1 transaction DAG constructed via the enabler-chain method.

Each enabler chain ends with a terminal enabler that connects to a \txname{WinPahse1} transaction. This transaction consumes the $1..N$ selector output, thus preventing any other party from winning. 

Depending on its position in the bracket’s player list, a registered asserter receives an
enabler chain that begins with one or more \emph{challenger} enablers and then continues
with \emph{asserter} enablers. The role for each pair $(X,Y)$ is fixed by the bracket:
the edge labeled $X/Y$ authorizes a dispute on $X$’s assertion with $Y$ as challenger.
For example, Bob first appears with a challenger enabler \textsf{A/B} (challenging Alice’s
assertion); if he advances, his next scheduled edge is \textsf{B/C}, where he acts as the
asserter against Charlie.

After a party $X$ registers for the first round, the subsequent \txname{EnableRound}$(i)$
transactions that advance $X$’s enabler chain are, by default, issued by $X$; to ensure
liveness, they are also pre-signed so that any honest participant may broadcast them on
$X$’s behalf. The tournament terminates when exactly one enabler chain reaches its
terminal link \textsf{Next $X(i)$ Enabler} and all competing chains have been cut via the designated penalization transactions, thereby disabling the remaining parties from further progression. There should be no race-condition to issue a \txname{WinPahse1} transaction. Watchtowers to monitor only the last winner selection transaction to learn which assertion they should evaluate. The last selector guarantees that all other Phase~2 templates remain dormant and cannot subsequently be activated. This wiring establishes a single, race-free handoff from Phase~1 to Phase~2.

\subsection{Stall Prevention}

If both paired asserters abstain and neither cuts the other, any third party may issue
a pre-signed \txname{DisputeTimout} transaction that performs a \emph{dual cut},
removing both from the bracket. 

We emphasize that separate stall-detection transactions are necessary: enforcing stalls via
multi-party, FLEX2-style components co-signed by all remaining asserters would inflate the signature burden to $\bigO(N^3)$.

In the worst case the honest asserter need issue at most one
\txname{DisputeTimout} transaction per round. 

\subsubsection{Soundness Proof for the Enablement Chains}
\label{app:echain}

We give an formal case analysis showing that the enablement scheme eliminates all but
one asserter and prevents post-loss participation.

\paragraph{Notation.}
In round $r$, let $E_A^{r}$ and $E_B^{r}$ denote Alice's and Bob's current enabler
inputs, and let $E_A^{r+1}$ and $E_B^{r+1}$ denote their next-link enablers. Transactions:
\txname{BobChallenge} opens the dispute; \txname{NoBobChallenge} is Alice's remedy if Bob
does not challenge; \txname{DisputeTimeout} is a stall detector.

\paragraph{Case 1: both enabled ($E_A^{r}$ and $E_B^{r}$ exist).}
\begin{enumerate}
  \item \emph{Bob challenges.} If \txname{BobChallenge} is issued and neither party
        finalizes nor penalizes the other, any third party may issue
        \txname{DisputeTimeout}, which consumes the winner selector and disables both from advancing.
  \item \emph{Bob does not challenge.}
    \begin{enumerate}
      \item \emph{Alice issues \txname{NoBobChallenge}.} This spend consumes
            $E_B^{r}$ and $E_B^{r+1}$; Alice should issue the next \txname{EnableRound} transaction to prevent the next contender issuing the stall detection transaction. 
      \item \emph{Alice does not issue \txname{NoBobChallenge}.} Any remaining
            party may issue \txname{DisputeTimeout}, which retires the dormant
            winner selection edge and disables both Alice and Bob from further participation in this
            bracket slot.
    \end{enumerate}
\end{enumerate}

\paragraph{Case 2: Alice enabled, Bob disabled ($E_A^{r}$ exists, $E_B^{r}$ absent).}
Alice waits until the winner selection timelock expires and issues the next \txname{EnableRound} transaction, advancing to the next round. Stall-detection is inapplicable since the winner selection output has already been consumed.

\paragraph{Case 3: Alice disabled, Bob enabled ($E_A^{r}$ absent, $E_B^{r}$ exists).}
Bob cannot proceed in the dispute: the \txname{BobChallenge} transaction requires Alice to be enabled. Bob  waits until the winner selection timelock expires and issues the next \txname{EnableRound} transaction, advancing to the next round. Stall-detection transaction is unavailable for the same input reasons as above.

\paragraph{Case 4: both disabled ($E_A^{r}$ and $E_B^{r}$ absent).}
No action is possible or required.

\paragraph{Conclusion.}
In each round, either (i) one party advances while the other's next-link is cut, or
(ii) both are removed by a stall detector. By induction over the rounds, exactly one asserter retains an unspent final enabler at the end of Phase~1; that party alone can initiate the Phase~2 refund request.

\subsubsection{Use of the FLEX Component}

\begin{figure}[h]
\centering
\includegraphics[width=0.7\textwidth]{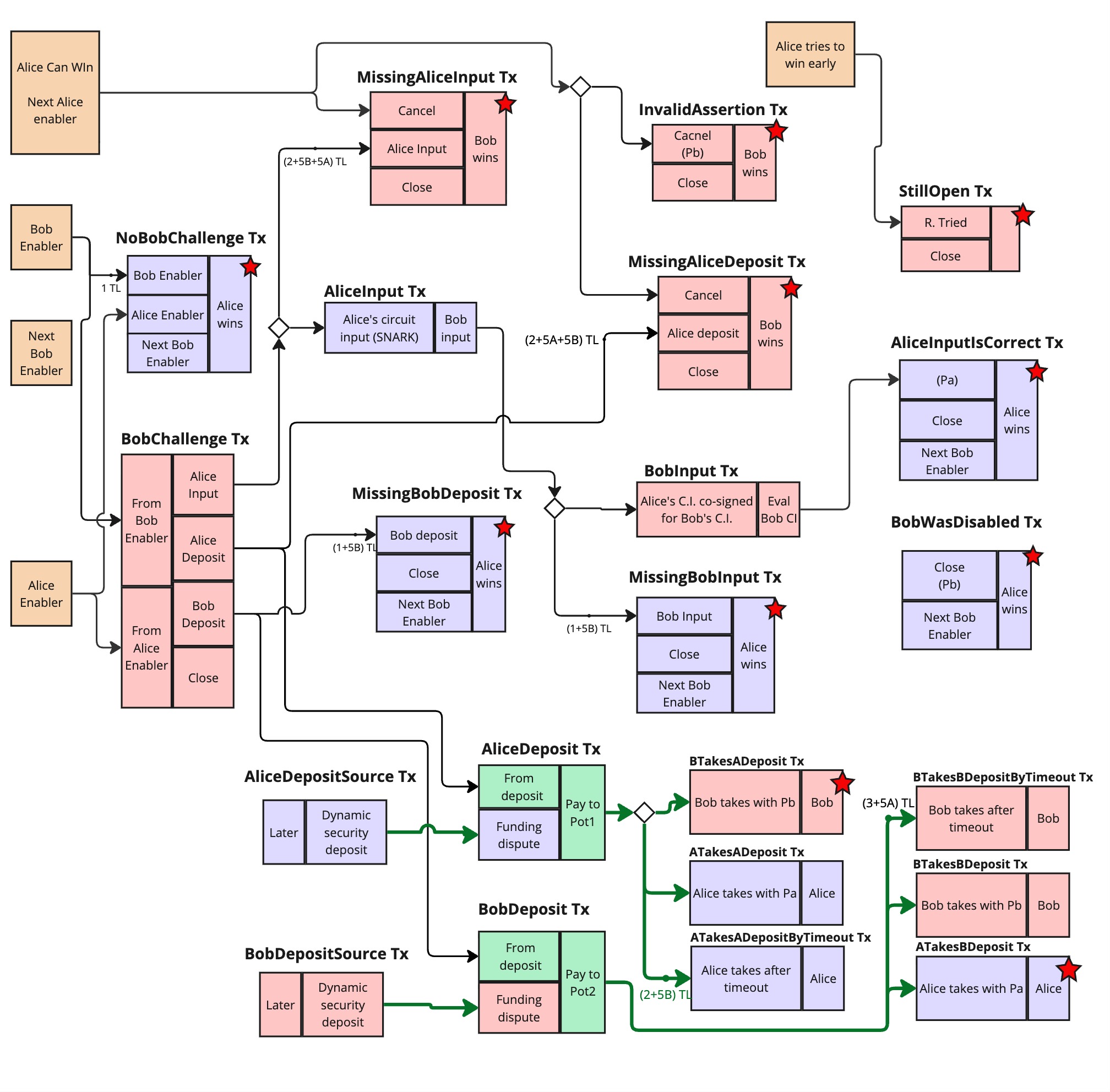}
\caption{A FLEX2 Component suitable for phase 1 and phase 2 of the BATTLE Tournament}
\label{fig:flex}
\end{figure}

To implement each tournament dispute, we instantiate an augmented \textsf{FLEX2} component
(Figure~\ref{fig:flex}) that supports \emph{chaining} across rounds. 

As in the original
\textsf{FLEX2}, exactly two parties interact—Alice (asserter) and Bob (challenger).
The augmentation adds (i) role-specific enabler inputs (for Alice’s assertion and Bob’s
challenge), (ii) a \emph{cut-next} spend that, upon a loss, allows the winner to disable
the loser’s next-link enabler.

In Figure~\ref{fig:flex}, transactions issued by Alice are shown in violet, those issued by Bob in pink, and transactions that may be broadcast by any participant (Alice, Bob, or a third party) in yellow. Orange boxes denote the input UTXOs consumed by the component. Green regions highlight the portions constrained by covenant signatures in partially signed transactions, i.e., fields whose values are fixed by pre-signed commitments while other fields remain free.

The \textsf{FLEX} component in Figure~\ref{fig:flex} consumes five input UTXOs:
\begin{enumerate}
  \item \textsf{Alice Can Win} (also serving as \textsf{Next Alice Enabler}),
  \item \textsf{Bob Enabler},
  \item \textsf{Alice Tries to Win Early},
  \item \textsf{Alice Enabler}, and
  \item \textsf{Next Bob Enabler}.
\end{enumerate}
Relative to the original \textsf{FLEX2}, the two additional inputs are
\textsf{Alice Enabler} and \textsf{Next Bob Enabler}. These enable \emph{chaining} across
rounds: any resolution in which Alice wins must spend the \textsf{Next Bob Enabler} input,
thereby cutting Bob’s subsequent enabler link; symmetrically, any resolution in which
Bob wins must spend the \textsf{Next Alice Enabler} input (aliased by \textsf{Alice Can Win}),
cutting Alice’s next link. 

Note that Figure~\ref{fig:flex} depicts generic transaction templates that show all
possible inputs for clarity. In an actual instantiation where no UTXO is wired to one
of the inputs (e.g., \textsf{Next Bob Enabler}), every transaction that would otherwise
consume that input \emph{must be constructed in a variant that omits it}: the input is
left out of the transaction’s \texttt{vin}, and the corresponding covenant commitments
and signatures are adjusted to the reduced input set. In other words, an unconnected
input does not appear on chain in the realized transactions.

For readability, the figure omits certain wiring edges (e.g., between \textsf{Next Bob Enabler}
inputs/outputs and \textsf{Close} inputs/outputs).

The meaning of each input is the following:

\textbf{Alice Can Win}. This is the output that allows Alice to win the dispute after it is closed. If Alice loses, this UTXO must be spent by the FLEX block.

\textbf{Bob Enabler}. This is the output that allows Bob to challenge and, therefore, to win. If this output is spent before the dispute starts, then Bob will be unable to challenge Alice.

\textbf{Alice tries to win early}. This is an output that Alice can dynamically make available during the progress of the dispute. Initially, this output is not present on the blockchain. As soon as it is available, a relative timelock allows Alice to win the tournament (i.e. be reimbursed in a bridge) irrespective of the state of the ongoing disputes. If this output becomes available for spending while a dispute is still ongoing, it is Bob’s responsibility to consume it as soon as possible to prevent misappropriation of funds. He does it by issuing the \txname{StillOpen} transaction. However, if this happens before Alice has issued an \txname{AliceInput} transaction containing the circuit input, then Bob will not be able to claim the funds locked as security deposits, and each party will take his own deposit after the timelocks expire.

\textbf{Alice Assert Enabler}. This is the output that allows Bob to challenge and, therefore, to win. If this output is spent before the dispute starts, then Bob will be unable to challenge Alice.

To set up the tournament for phase 1, the FLEX inputs are connected from different outputs. The 'Alice Can Win' FLEX input is repurposed as the 'Alice next enabler', but the 'Alice tries to win early' FLEX input is not used in this phase.

The input 'Alice Enabler' and 'Alice Next Enabler FLEX' comes from Alice's chain of enablers, and so are Bob’s enablement inputs taken from Bob’s chain.

The FLEX block has two parameters, $A$ and $B$, which represent the maximum delay that Alice and Bob can incur, measured in epochs, before making the security deposit. Since Alice is second to post the deposit, Alice's delay starts counting after Bob’s delay. For example, if $A=0$ and $B=1$, Bob will have one epoch to post the deposit, and Alice will need to do the same immediately afterward. Because we do not use the optimistic early win method of FLEX, we set $A=0$ and $B=0$ in this phase. To use the optimistic early win method, we would set staggered values for $B$ in each round.

An informal proof of the correct construction of the enablement chain can be found in Appendix~\ref{app:echain}.

\subsubsection{Tournament}

\begin{figure}[h]
\centering
\includegraphics[width=0.7\textwidth]{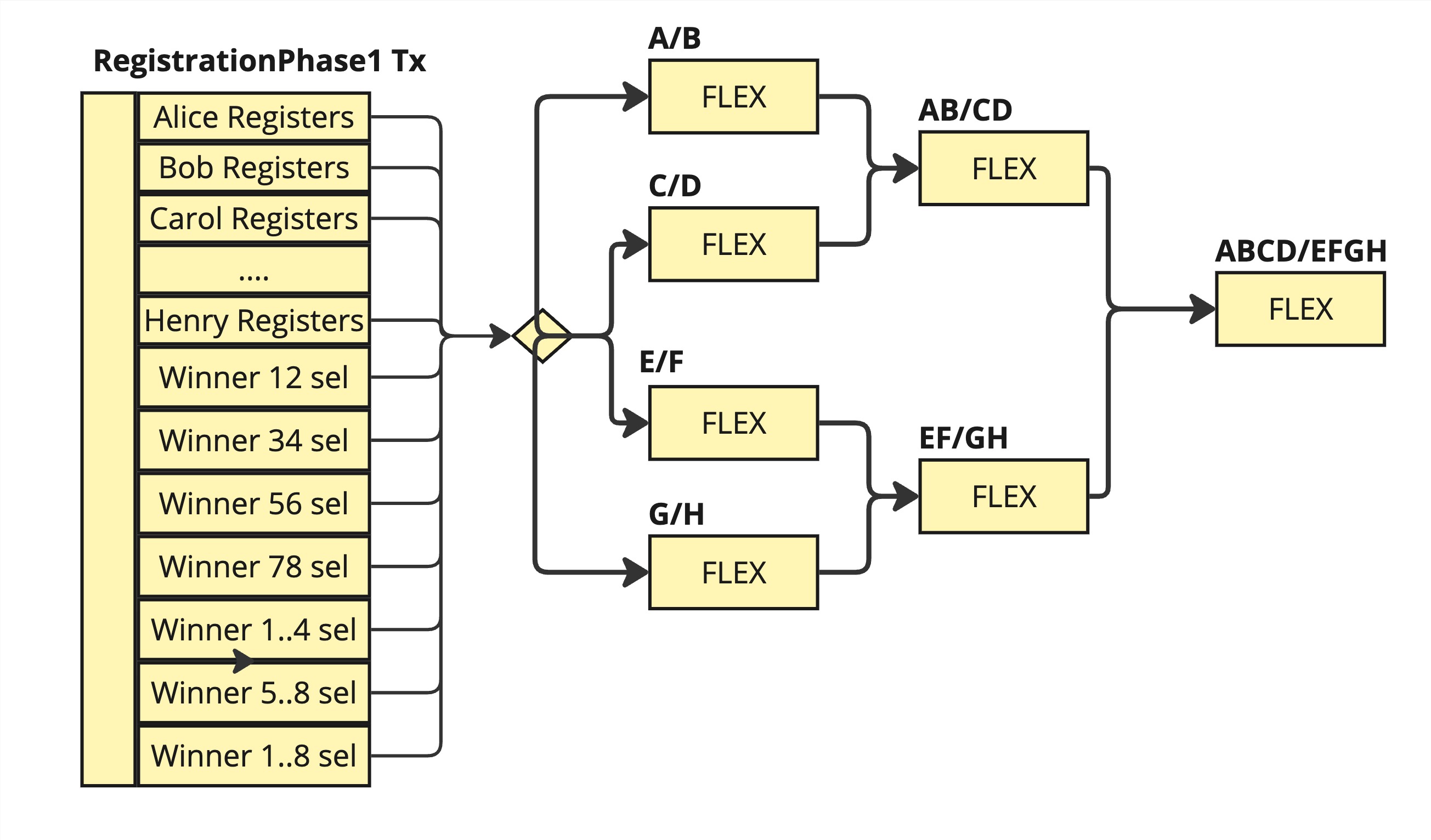}
\caption{A Phase 1 Tournament transaction DAG for 8 asserters, with all FLEX components in each round superimposed. Tick lines represent superimposed connections}
\label{fig:phase1}
\end{figure}

Phase~1 is a single-elimination (binary) bracket: each scheduled dispute removes one
participant, and the winner advances while redeeming a bounty \(\ADR\) taken from the
loser’s on-demand bond \(\AOSB\). Figure~\ref{fig:phase1} illustrates the bracket after
registration for eight asserters; pairings are fixed by the registration order and
are activated round-by-round via timelocks.

Some \textsf{FLEX} boxes in the diagram are \emph{superimposed} placeholders for multiple
pairwise components. For example, the box labeled \texttt{ABCD/EFGH} stands for all
$4\times 4=16$ pairings between $\{A,B,C,D\}$ and $\{E,F,G,H\}$—namely
$A/E, A/F, A/G, A/H, B/E, B/F, \ldots, D/G, D/H$. This grouping explains why the
Phase~1 construction requires pre-creating $\bigO(N^{2})$ transaction templates and signatures.

\begin{figure}[h]
\centering
\includegraphics[width=0.7\textwidth]{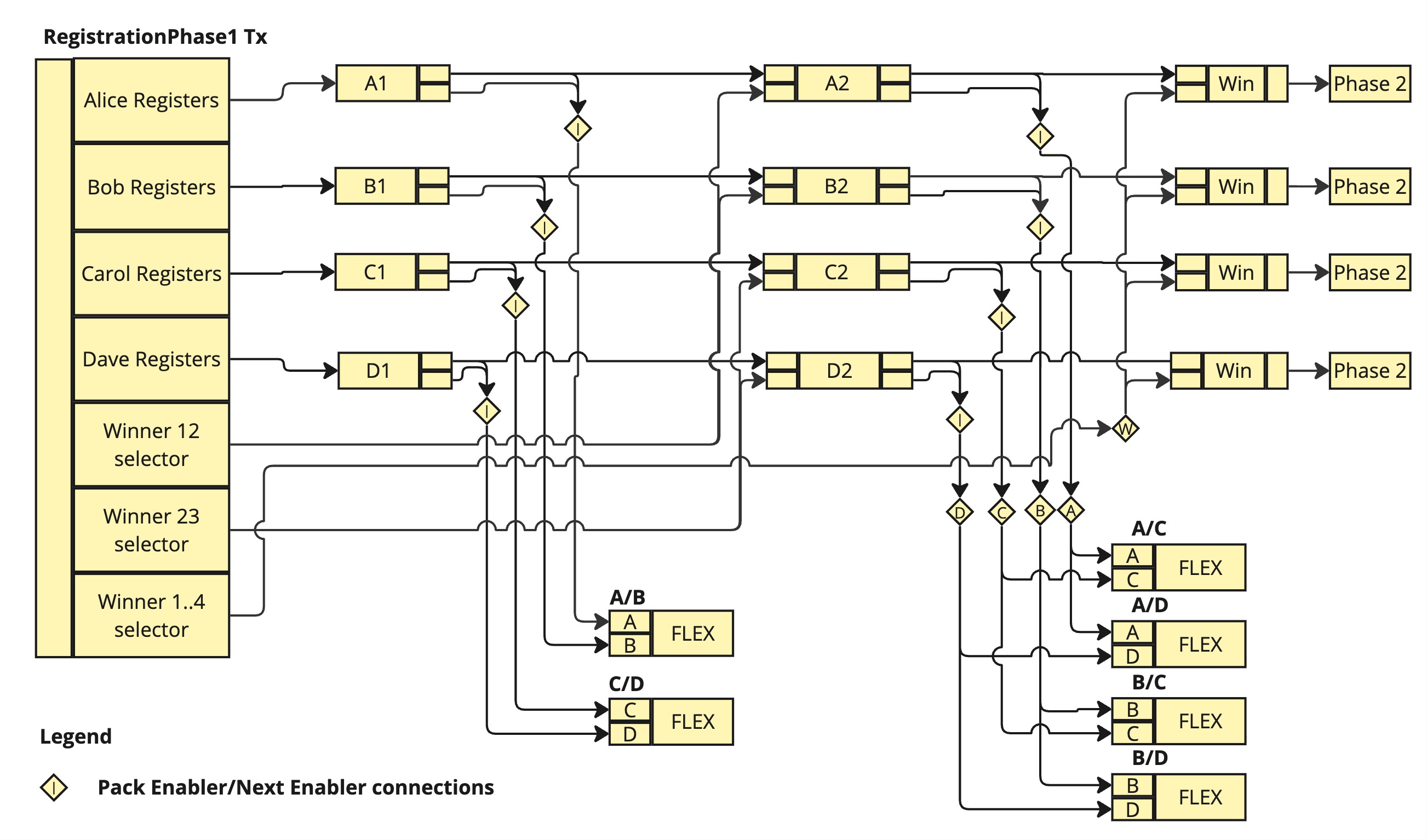}
\caption{A Phase 1 Tournament transaction DAG for 4 asserters with their enablement chains, with all precreated FLEX components shown separately}
\label{fig:phase1-4}
\end{figure}

Figure~\ref{fig:phase1-4} depicts the \textsf{FLEX} components for a four-asserter
tournament over two rounds. In the second round, only the single component corresponding
to the actual finalist pairing is instantiated on chain; the alternative component
remains dormant and is never broadcast.

To instantiate a \textsf{FLEX} component for a specific bracket edge, map the role
placeholders to the concrete parties: \emph{Alice} denotes the \emph{asserter} and
\emph{Bob} denotes the \emph{challenger}. Thus, for a pairing $(X,Y)$ with $X$ acting
as asserter and $Y$ as challenger, set \textsf{Alice}$\gets X$, \textsf{Bob}$\gets Y$
and wire the inputs/outputs (\textsf{Alice Enabler}, \textsf{Bob Enabler},
\textsf{Next Alice Enabler}, \textsf{Next Bob Enabler}) to the corresponding UTXOs.
If the bracket assigns the opposite roles, apply the symmetric mapping
(\textsf{Alice}$\gets Y$, \textsf{Bob}$\gets X$).

\subsubsection{The Tournament Chain (TC)}

We require a mechanism that prevents \emph{concurrent} tournaments for the same
peg-in slot while still permitting \emph{sequential} assertions in arbitrary order.
A naive approach that penalizes any party opening a tournament on a conflicting
assertion after another is already active is vulnerable to race conditions: two
parties may attempt to open concurrently without knowledge of each other, and such
benign concurrency should not incur penalties.

We propose the \emph{Tournament Chain} (TC), a transaction chain that exposes
rate-limited \emph{slots} for opening tournaments. Each slot becomes available
according to preset timelocks; opening a tournament consumes the next available
slot and anchors its tournament instance. If two openings race for the same slot,
only the first confirmed spend binds the slot, and the conflicting attempt cleanly
fails without penalty. The TC thus enforces bounded concurrency (at most one live
tournament per slot) and provides a deterministic admission schedule for sequential
assertions, while remaining race-free and non-punitive under simultaneous attempts.

\begin{definition}[Tournament Chain (TC)]\label{def:TC}
The \emph{Tournament Chain} (TC) is an on-chain, singly linked sequence of transactions
\[
\txname{TCStart} \;\to\; \txname{OpenTournament}_1 \;\to\; \txname{OpenTournament}_2 \;\to\; \cdots
\]
where each link $\txname{OpenTournament}_{i+1}$ spends a designated \emph{next-link} output of
$\txname{OpenTournament}_i$ and is encumbered by a relative timelock $t$ (or $t_i$) that rate-limits
issuance. Each $\txname{OpenTournament}_i$ exposes a distinct output \txname{StartPhase1}; spending
this output instantiates a Phase~1 tournament for that \emph{slot}. At most one tournament can be
anchored per slot: concurrent attempts to spend the same \txname{StartPhase1} are resolved by normal
blockchain ordering, with the first confirmed spend binding the slot and conflicting spends failing
without penalty.
\end{definition}

Any party may advance the TC by publishing the next \txname{OpenTournament} link once its
relative timelock elapses, thereby enabling a new tournament slot. The parameter $t$
sets the minimum inter-slot interval and thus the maximum admission rate: at most one
tournament can be opened every $t$ timelock periods. For example, choosing
$t = 2 + 2\lceil \log_2 N \rceil$ (two registration windows plus the worst-case
$\lceil \log_2 N \rceil$ Phase~1/2 rounds) ensures that tournaments do not overlap.

Figure ~\ref{fig:tc} illustrates this simple scheme:

\begin{figure}[h]
\centering
\includegraphics[width=0.7\textwidth]{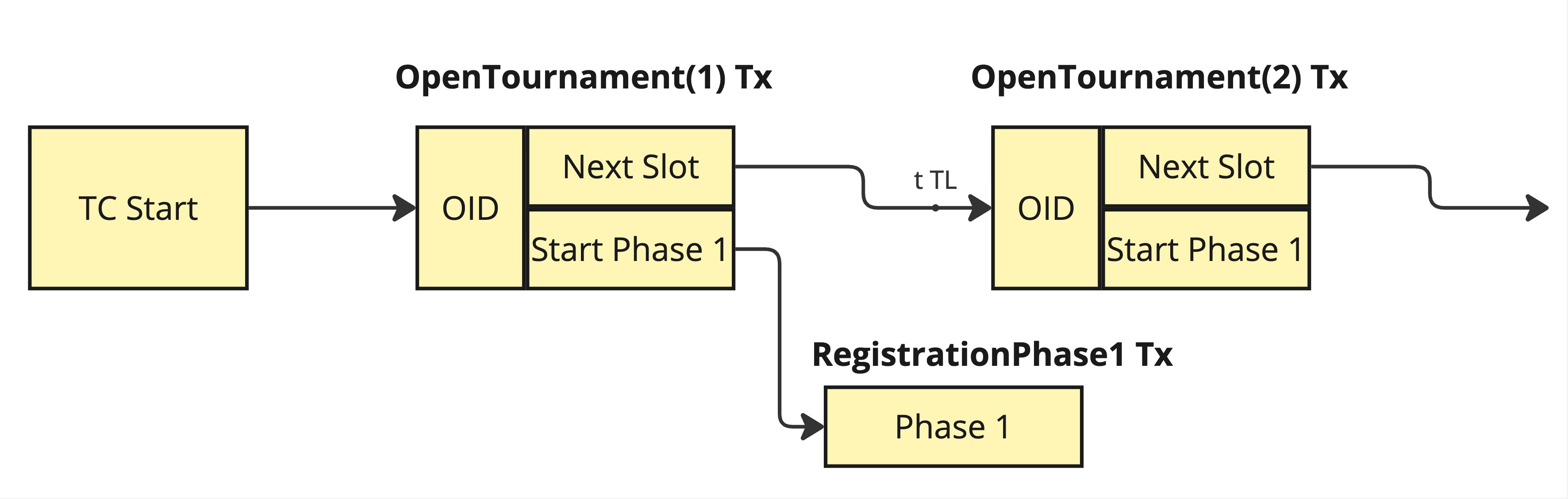}
\caption{The tournament chain that allows tournaments to occur synchronously without overlaps}
\label{fig:tc}
\end{figure}

Controlled concurrency can be increased by parameterization: (i) reduce the inter-link
timelock $t$ to admit slots more frequently; (ii) expose $m>1$ \txname{StartPhase1}
outputs per \txname{OpenTournament} link (up to $m$ concurrent tournaments per slot);
or (iii) deploy $K$ parallel TCs keyed to disjoint namespaces (e.g., outcome classes).
Each method raises throughput while preserving the race-freedom of per-slot openings;
the chosen parameters must still respect fee and liveness bounds.

\subsubsection{The Open-and-Abandon Tournament Attack (O\&A)}

A vulnerability of the TC is that a party can consume a slot by broadcasting
\txname{OpenTournament} but then refrain from posting any assertion during the
associated registration window. If no other party registers a conflicting assertion
for that tournament, the pre-signed tournament sub-DAG anchored at that TC link is
wasted: all signatures are bound to the link’s transaction identifier, so the
sub-DAG cannot be reused and must be re-created. We refer to this as the
\emph{Open-and-Abandon} (O\&A) attack.

\begin{definition}[Open-and-Abandon tournament attack (O\&A)]\label{def:QAA}
An \emph{Open-and-Abandon} (O\&A) attack occurs when an adversarial party $u$
broadcasts an \txname{OpenTournament} transaction for a TC slot and the
corresponding registration window elapses with \emph{no} assertion posted for
that tournament (by $u$ or any other party).
\end{definition}

To mitigate O\&A, each \txname{OpenTournament} must carry the publisher’s operator
identifier \(\mathsf{OID}\), making the slot consumption attributable. If a party opens
a tournament and then fails to register an assertion, the side-system enforces a penalty:
its persistent bond \APSB\ is slashed, the participant is scheduled for removal, and the
affected funds are migrated to a fresh TC instance.

This mitigation entails pre-creating and co-signing \(\bigO(WN^{2})\) transaction templates,
where \(N\) is the number of operators and \(W\) is the number of TC links provisioned
as buffer to remain operational during ongoing O\&A. Concretely, \(W\) should at least
cover the number of active bridge UTXOs that may require migration before the current
TC becomes unusable, plus additional links sufficient to bridge the period until an
operator-removal/migration decision is finalized (whether by contract logic or by
social consensus).
 
A complementary mitigation is to tie the TC advancement rate to the number of committee
signatures collected on each link. Let $N$ be the committee size. The \textsf{NextSlot}
output of every \txname{OpenTournament} pays to a Taproot whose TapTree contains, for
each $i\in\{1,\ldots,N\}$, a leaf that: (i) verifies an $i$-of-$N$ committee signature,
and (ii) enforces a relative timelock of $T_Z/i$ units. Consequently, the inter-link
delay is inversely proportional to the number of signers: with $i$ signatures, the
next TC link can be issued after $T_Z/i$ units, yielding a maximum delay $T_Z$ (for
$i=1$) and a minimum delay $T_Z/N$ (for $i=N$). A single malicious party cannot
unilaterally accelerate issuance and is rate-limited to one link per $T_Z$, whereas a
broadly cooperative committee can admit new slots at up to $N$ times that rate.

A further mitigation is to require an on-demand security bond \AOSB\ to open a tournament.
However, under FLEX-style bonding the funding transaction that creates the \AOSB\ UTXO
has a party-specific \texttt{txid}. In the absence of \txname{SIGHASH\_NOINPUT}
(\emph{a.k.a.} \txname{ANYPREVOUT}), the spending paths in a shared pre-signed DAG
cannot be made agnostic to that unknown funding \texttt{txid}. Consequently, a single
tournament DAG cannot accept an \AOSB\ supplied by an arbitrary opener. Emulating this
flexibility would require maintaining $N$ parallel DAG variants (one per would-be
opener/funding path), multiplying the number of pre-signed tournament DAGs by $N$.

Identifying more efficient mutual-exclusion primitives, designing opener-agnostic on-demand security bonds for tournament admission, and developing sharper, verifiable rate limiters for tournament openings remain open problems for future work.

\subsection{Phase~2}

In Phase~2, the surviving assertion is open to challenge by any party. The Phase~2 Transaction DAG is designed as $N$ mutually exclusive tournament templates, one for each party
that could act as the Phase~2 asserter, but only a single template will be activated on
chain.

The first output of the unique \txname{WinPhase1} issued serves as the activation input for the Phase~2 template of the surviving asserter; all other Phase~2 \txname{WinPhase1} transaction
cannot subsequently be activated, since the winner selection output (in the  \txname{RegistrationPhase1} transaction) can only be consumed once.

Although we do not include it in the presented design, it is possible to prevent losing asserters from Phase~1 from challenging at Phase~2 by making Phase~1 penalization transactions contain output that, in conjuntion with Phase 1-2 cancellation transaction, consume the enablers of those asserters in the roles of Phase~2 challengers.
 
\subsubsection{Registration}

A single timelock period is allocated for challenger registration. If no challenger
registers, the surviving assertion is accepted. To register, a challenger must issue the \txname{RegInPhase1} transaction, and later deposit the security bond of the FLEX component it should take part. If a malicious party tries to register after one timelock has elapsed, the asserter  can issue the \txname{RegTimeout} transaction and cancel the registration attempt. The honest asserter’s transaction cost is independent of the number of honest parties. If all parties are honest, the asserter can attempt an early refund or simply the asserter can wait until the Phase~2 reaches its maximum tournament time, and issue the \txname{Refund} transaction.

Registered challengers are ordered by a uniform random permutation fixed at setup for the corresponding Tournament Chain (TC) link; each TC link is assigned an independent permutation, yielding a static schedule per link.

The asserter may initiate any eligible dispute immediately—
either because a challenger has already timed out or because sufficient capital is
available to run multiple disputes in parallel—but may not defer any dispute beyond
its scheduled deadline. The asserter may also preemptively initiate a dispute to realize its reward earlier and
thereby finance multiple subsequent disputes.

\subsubsection{Transaction DAG}

\begin{figure}[h]
\centering
\includegraphics[width=0.7\textwidth]{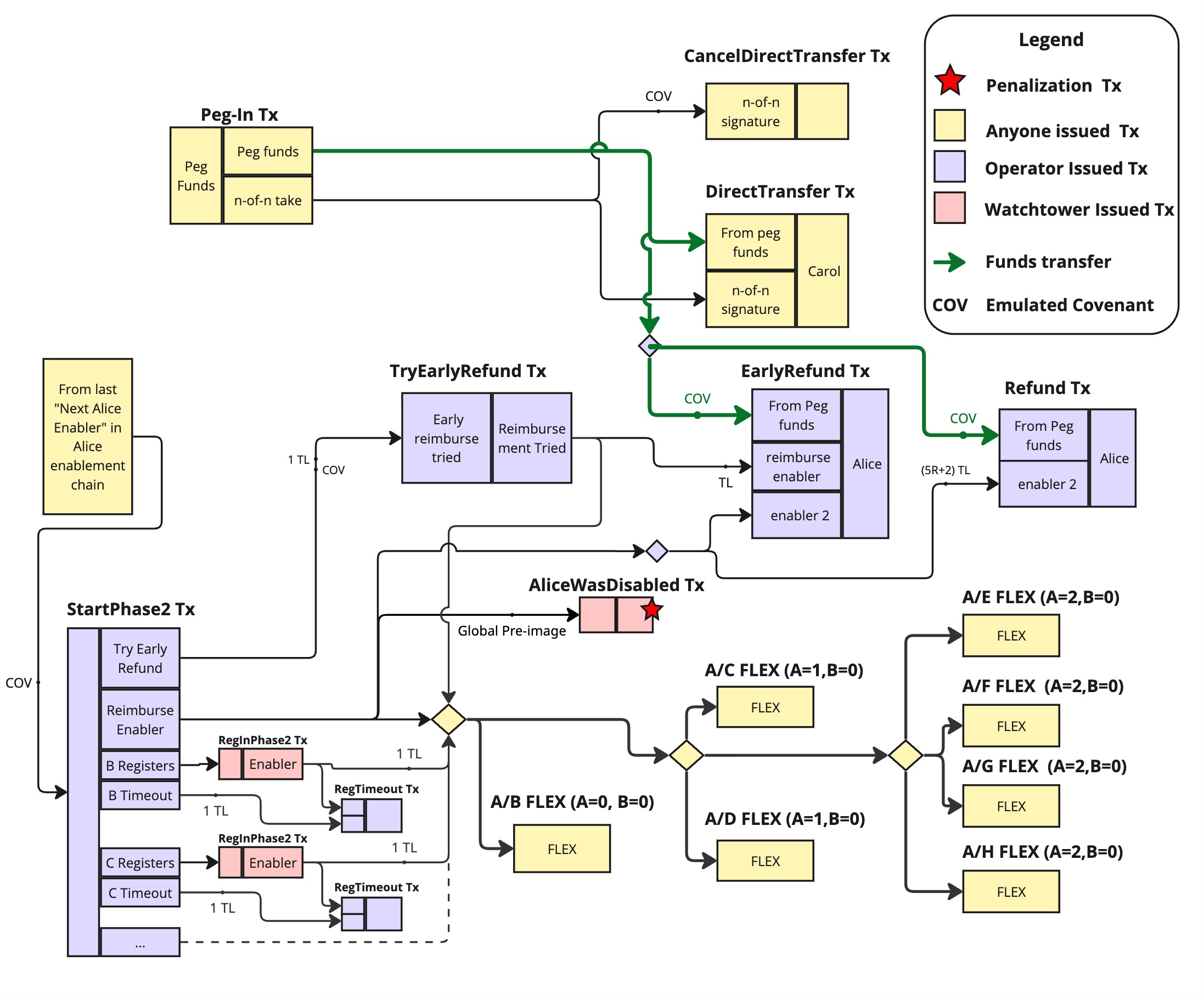}
\caption{The transaction DAG in a single peg-in / single operator, BATTLE Bitcoin Bridge}
\label{fig:phase2}
\end{figure}

We implement BATTLE Phase~2 on Bitcoin. Figure~\ref{fig:phase2} depicts the transaction
DAG for a FLEX-based bridge with a single peg-in, where one operator (Alice) serves as
the Phase~2 asserter against seven registered challengers $B,\ldots,H$. In this setting,
the operator fronts the funds and is reimbursed upon successful dispute resolution.

When the operator has won all the disputes, he can issue the \txname{TryEarlyRefund}/ \txname{EarlyRefund} combo or simply the \txname{Refund} transactions. The early refund mechanism is a two-step process that allows the operator to be quickly reimbursed if very few parties challenge (even before $\log(C)$ tournament rounds) or if he has the capital to simultaneously engage and finalize all disputes in a single round. Each dispute is handled by a FLEX component with a variable delay parameter for Alice ($A$) and a zero delay for Bob ($B=0$). A delay of $A=0$ implies that Alice's security deposit in that component cannot be delayed. Alice's security deposit can be delayed for $5A$ timelock periods, which is equivalent to one epoch.

Note that Alice can choose when to post the security bonds for all challengers, but each challenger has a different deadline to do so. 

For the disputes in Phase~2 we also use the FLEX component previously presented, but we connect it differently. The 'Alice Can Win' FLEX input comes from the 'Reimburse Enabler' output. The 'Bob Enabler' comes for the enabler with the same name in the \txname{StartTournament} transaction. The 'Alice tries to win early' FLEX input comes from the Reimbursement Tried output in the \txname{TryEarlyRefund} transaction.
 
Alice uses the \txname{TryEarlyRefund} transaction to get refunded if all disputes are over before the ($5R+2$) timelock periods have elapsed, which represents the deadline of the tournament. The \txname{StillOpen} transaction penalizes Alice if she tries to be reimbursed with the \txname{TryEarlyRefund} transaction while at least one dispute is not over.

To avoid race conditions, Alice should only issue \txname{TryEarlyRefund} if all open disputes on that round have been closed and all unopened disputes have been canceled. Disputes are canceled by issuing the corresponding \txname{NoBobChallenge} transactions in the FLEX blocks. 

If there are too many (i.e. thousands) of inactive challengers and Alice does not want to cancel each one of them, then Alice can wait until all predefined rounds in the tournament are over and issue the \txname{Refund} transaction.

\begin{figure}[h]
\centering
\includegraphics[width=0.7\textwidth]{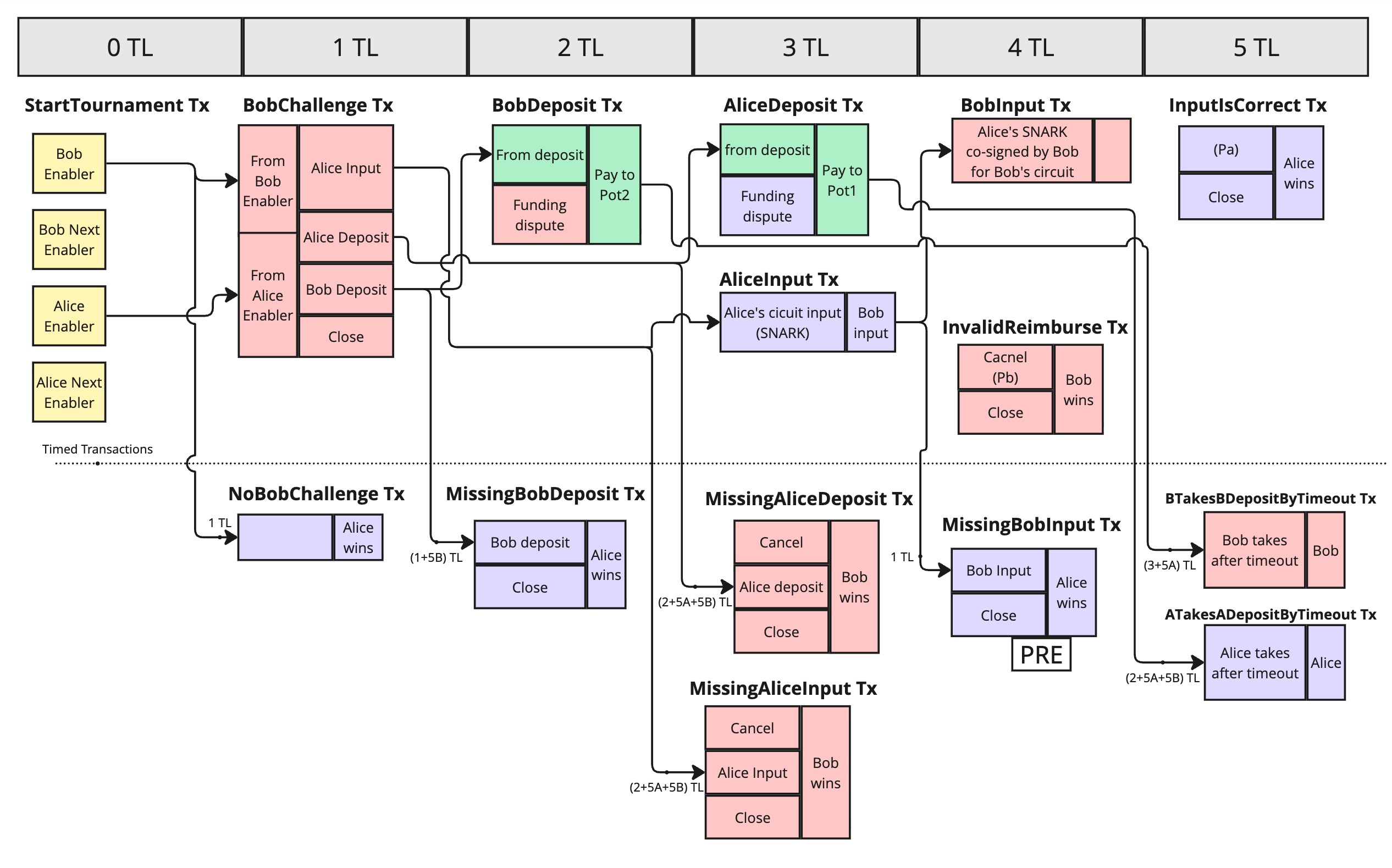}
\caption{The worst-case delay for the publication of the FLEX transactions. The transactions in the bottom half restrict the ones in the upper half.}
\label{fig:delay}
\end{figure}

Figure~\ref{fig:delay} depicts the worst-case timeline for a single \textsf{FLEX} block,
measured in timelock (TL) periods, under the assumption that each party broadcasts every
required transaction at the latest admissible moment without incurring penalties.

\begin{figure}[h]
\centering
\includegraphics[width=0.7\textwidth]{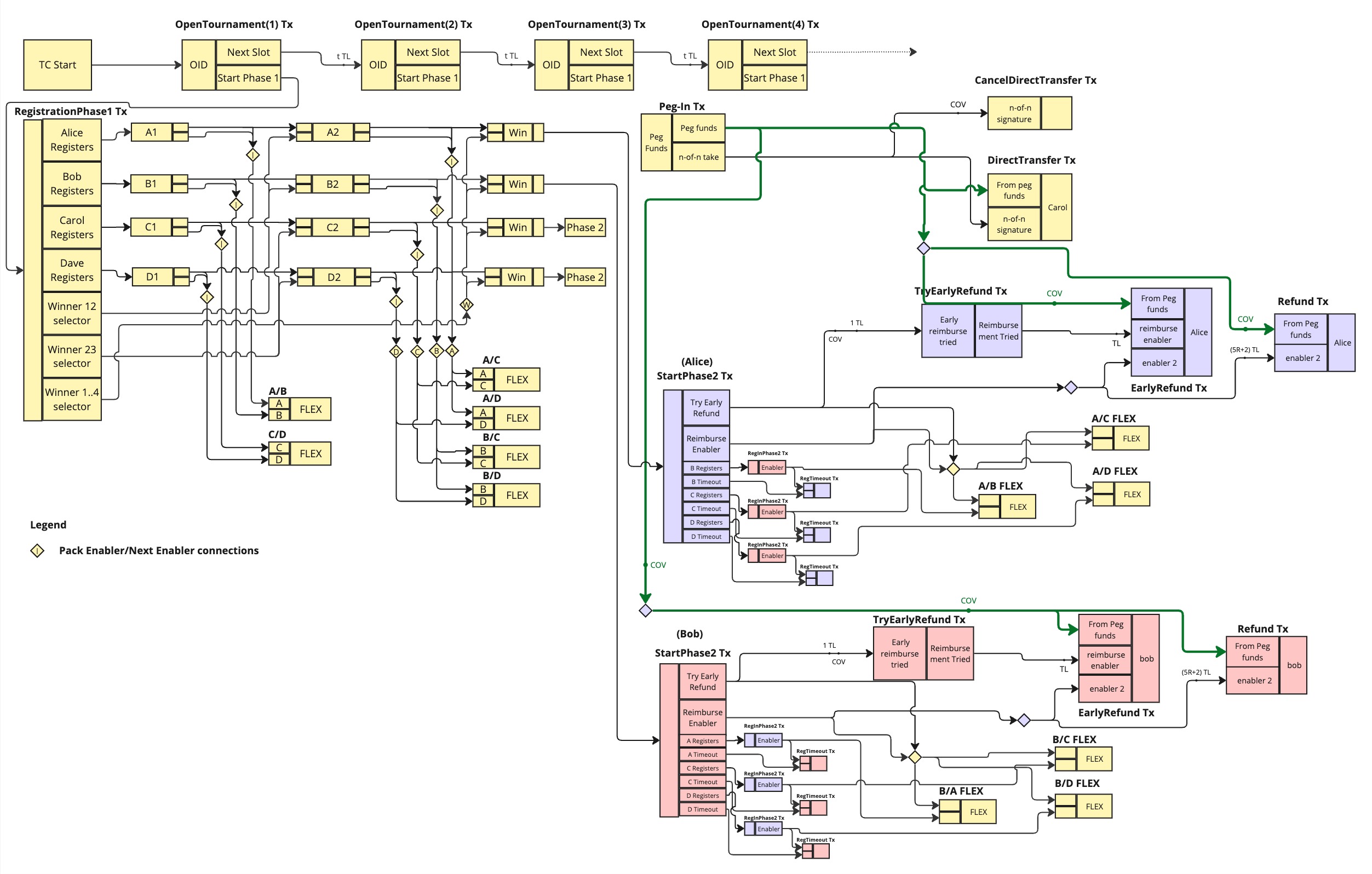}
\caption{Putting all the pieces together: A tournament chain, shown with a Phase 1 and two Phase 2 tournaments, for $N=4$}
\label{fig:all-together}
\end{figure}

Figure~\ref{fig:all-together} presents the end-to-end transaction DAG for a four-participant
instance of BATTLE. It composes the Tournament Chain (TC), a single TC slot with its Phase~1
bracket, and the two Phase~2 tournament templates corresponding to the possible Phase~1
enabler emitted by the Phase~1 winner.

\subsection{Extensions}

\subsubsection{Disabling Losing Parties Using Global Secrets}

Because on-demand security bonds (\AOSB) are posted per dispute and settled independently, a
losing party need not be universally disabled to protect capital flows. However, it can be
desirable to bar a loser from participating in sub-protocols that do not require \AOSB\
(e.g., committee voting). To this end, each party $u$ commits at setup to a \emph{global
disable secret} by publishing $c_u=H(s_u)$. When $u$ loses a dispute, the winner obtains $s_u$ from the garbled circuit and  reveals $s_u$ on chain. After revelation, any attempt by $u$ to assert or
challenge matches a “disable” branch that immediately blocks the action via the pre-signed
\txname{AliceWasDisabled} transaction in the main DAG (or the \txname{BobWasDisabled}
transaction within the \textsf{FLEX} block).

We describe mechanisms to instantiate the per-party \emph{disable secret} without the need to add more output labels to each garble circuit to reveal $c_u$. For each ordered
pair $(X,Y)$, let $P_{X,Y}$ be the pairwise preimage that $X$’s circuit reveals to $Y$
if $X$ loses to $Y$, and let $H_{X,Y}\!=\!H(P_{X,Y})$ be its setup commitment. Each party
$X$ also commits to a global disable secret $G_X$ via $H_X\!=\!H(G_X)$.

\paragraph{Method~1 (direct reveal in GC).}
Every garbled circuit authored by $X$ is wired so that, on any losing transcript, it
outputs $G_X$. In Yao’s garbling, $X$ controls output labels; $X$ proves in zero
knowledge at setup that a single-bit circuit’s losing output label decode to a value hashing to $H_X$.
When $X$ loses, $G_X$ is revealed on chain and can drive the “disable” branches.

\paragraph{Method~2 (encryption under pairwise keys).}
Let $\mathsf{Enc}_K(\cdot)$ and $\mathsf{Dec}_K(\cdot)$ be a fixed symmetric cipher with
$\mathsf{Dec}_K(\mathsf{Enc}_K(M))=M$. There are two variants:

\emph{(2a) Single-loss disable.} For each counterparty $Y$, $X$ publishes
\[
E_{X,Y} \;=\; \mathsf{Enc}_{P_{X,Y}}(G_X),
\]
and provides a zero-knowledge proof that it knows $(P_{X,Y},G_X)$ such that
$H(P_{X,Y})=H_{X,Y}$, $H(G_X)=H_X$, and $\mathsf{Enc}_{P_{X,Y}}(G_X)=E_{X,Y}$.
Upon losing to $Y$, revealing $P_{X,Y}$ lets any party recover $G_X=\mathsf{Dec}_{P_{X,Y}}(E_{X,Y})$.

\emph{(2b) Threshold disable.} To soften penalties and tolerate errors, $X$ secret-shares
$G_X$ into $m$ shares $\{S_{X,j}\}$ with threshold $t$ (e.g., Shamir). For each designated
counterparty $Y_j$, $X$ publishes $E_{X,j}=\mathsf{Enc}_{P_{X,Y_j}}(S_{X,j})$ and proves
in zero knowledge that $H(P_{X,Y_j})=H_{X,Y_j}$ and that the shares reconstruct to a value
hashing to $H_X$. Only after $X$ loses to at least $t$ distinct parties (revealing the
corresponding $P_{X,Y_j}$) can the decrypted shares be combined to reconstruct $G_X$.

In both methods, the disable secret (or enough shares thereof) is revealed only after
on-chain evidence of loss, and all statements about $H_{X,Y}$ and $H_X$ are enforced via
setup-time zero-knowledge proofs without exposing $P_{X,Y}$ or $G_X$ prematurely.

\subsubsection{Single Circuit Input Publication}

The dominant per–tournament costs are (i) capital temporarily locked in on–demand
security bonds and (ii) on–chain publication of the circuit input. In some bridge deployments
there is also a separate capital cost: liquidity fronted to users until reimbursement
completes. Both capital components scale as $O(\log N)$ in the worst case (by the round
structure), while the fronting cost additionally depends on the amount advanced. Although TOOP~\cite{futoransky2025toop} can eliminate fronting costs, it does not meet our scalability requirements, as it supports only a limited number of operators. In what
follows we disregard the fronting cost and focus on the two costs inherent to BATTLE:
input publication and on–demand security bonds. 

\begin{example}
Absent input-compression mechanisms (e.g., WISCH~\cite{wisch}), the circuit input must be carried via
Lamport signatures. A typical token-burn SNARK proof is \(\sim128\) bytes. Lamport
verification in Bitcoin Script consumes \(\sim400\) witness bytes per signed byte, i.e.,
\(\sim128\times 400 \approx 51\,\mathrm{KB}\) of witness per party per dispute. At
Sep.~2025 fee levels this is \(\sim\$160\). An optimized DV–SNARK as short as \(32\)
bytes would reduce publication cost to \(\sim12.8\,\mathrm{KB}\), i.e., \(\sim\$40\).
\end{example}

The per-dispute input-publication expense lower-bounds the on-demand security bond
\AOSB: to keep winners whole and enable reward recycling, \AOSB\ must cover (at least)
the worst-case L1 publication (and associated fee) for a single dispute. Consequently,
if publication cost is reduced by a multiplicative factor $x$, the same \AOSB\ can
underwrite up to $x$ simultaneous disputes, shortening the total tournament duration
and, indirectly, the time during which fronted liquidity remains outstanding. 

Reducing input-publication cost is therefore critical. A natural idea is to let the
asserter publish a \emph{single} input that serves all pairwise garbled circuits with
potential challengers. This is problematic: some or all challengers may abstain, and,
once input labels are revealed, those circuits cannot be safely reused, inflating the
number of garbled circuits by the number of potential peg-outs.

An alternative is to provision, per peg-out $u$, a set of \emph{neutral} input labels that
do not unlock any circuit by themselves, and to disclose, per dispute, a secret \textit{conversion key}
$K_{XY,u}$ that transforms the neutral labels of pegout $u$ into the specific labels for the $(X,Y)$
pair (e.g., by XOR-ing each neutral input label with $K_{XY,u}$). This avoids publishing
functional labels up front but shifts cost to setup: neutral label sets must be distinct
per peg-out, and each party must generate one-time Lamport keys per peg-out slot. For
$U$ peg-outs and input size $B$ bits each party must share
$U \cdot 2B $ Lamport public keys and $U$ hashes of conversion keys with the counterparties. A rigorous evaluation of this approach is left to future work.

\subsubsection{Contestable vs.\ Non-Contestable \textsf{FLEX} Components}

It is desirable to minimize the number of confirmation blocks used to attest chain
events: fewer confirmations improve user experience and reduce prover costs. For
instance, with attacker hashrate fraction $q=0.1$, a payment with $k=3$ Bitcoin
confirmations is reversed with probability on the order of $1\%$ under the classical
Nakamoto analysis~\cite{bitcoin_whitepaper}. However, such tail probabilities do not
capture \emph{expected profit}, which scales with the value at risk and, crucially,
with the aggregation of \emph{simultaneous} payouts that a single reorganization could
double-spend.

A payee can raise $k$ with the payment amount, but cannot generally observe whether
the payer is concurrently paying others. Correlated payouts increase the attacker’s
reward for a single reorg, invalidating fixed-$k$ assurances calibrated to a lone
transfer. Consequently, stateless SPV-style attestations that treat finality as a
function only of local headers and a fixed confirmation count are insecure unless
the protocol designers can upper-bound the adversary’s budget and the aggregate
value exposed, and then choose $k$ against the \emph{worst} feasible attack.

This motivates \emph{contestable} \textsf{FLEX} components: the asserter’s proof of an
event (e.g., inclusion on a given branch) is admissible, but challengers may submit
counter-proofs (e.g., a heavier/longer canonical branch or a conflicting event), and
the dispute circuit selects the winner by an objective fork-choice rule. Non-contestable
components (which accept an assertion after a fixed $k$ without counter-evidence) are
only appropriate under strong external bounds on exposure; otherwise, contestability
is required to maintain security under unknown, time-varying incentives.

An alternative assumption is that beyond a sufficiently large confirmation depth
(e.g., $k=144$ Bitcoin blocks), rational attackers will abstain because a successful
reorganization would depress the BTC price and erode realized profits. Both this
assumption (a hard bound on reorg depth) and the budget-bound assumption force smart
contracts to be parameterized for worst-case scenarios, which translates into large
$k$ and long delays, degrading user experience.

Accordingly, well-designed optimistic Bitcoin bridges admit \emph{contestable proofs}:
a canonical asserter submits a chain proof of the claimed event, and challengers may
submit counter-proofs for chains with strictly higher cumulative difficulty that do
\emph{not} contain the event~\cite{back2014sidechains}. The dispute circuit applies
the fork-choice rule to determine the winner. We refer to such designs as
\emph{contestable proof protocols}.

We present two constructions for \emph{contestable} \textsf{FLEX} components within
BATTLE.

\paragraph{Method~A (dual-proof input).}
Bob appends his counter-proof in \txname{BobInput}. Alice then co-signs the
counter-proof via a new \txname{AliceInputCoSig} transaction, ensuring authorization and binding
to the same dispute instance. Both the asserter’s proof and the challenger’s
counter-proof are provided as inputs to \emph{both} garbled circuits; the circuits
verify both and select the winner under the fork-choice rule (e.g., longest/heaviest
chain). The drawback is that, in the worst case, the circuit must include two
verifiers, approximately doubling its size relative to a non-contestable design.

\paragraph{Method~B (score-carry with single-proof verification).}
When the fork choice is determined by a scalar \emph{score} (e.g., cumulative
difficulty), we avoid dual verification. Let $C_A$ and $C_B$ be Alice and Bob chains. Let $S_A$ and $S_B$ be the scores of Alice and Bob chains, respectively. Let PegOutID be the peg out unique identification number and PegOutPos a locator of the transaction containing the pegout event in the L2 blockchain (e.g. the block number and the transaction index). Alice creates the proof $SNARK_A$ while Bob creates the proof $SNARK_B$. We use two one-time signature schemes: Winternitz for transferring values to scripts and Lamport for transferring values to scripts and for garbled circuit inputs. 

\begin{enumerate}
  \item Alice publishes publishes PegOutID, PegOutPos+ $S_A$, $SNARK_A$,  Lamport signed as part of her transaction \txname{AliceInput}. Her circuit verifies the $SNARK_A$ which enforces that (i) the pegout event exists in the blockchain $C_A$ at position PegOutPos, (ii) the event is associated with the ID PegOutID,  (iii) $C_A$ score is $S_A$ and (iv) there is $w$ work past the pegout event, for a predetermined $w$.

  \item Bob publishes his own $SNARK_B$, together with $S_B$, PegOutID, PegOutPos, co-signed with Lamport, together with Alice's signatures of PegOutID, PegOutPos so that the script can guarantee Bob is copying Alice's values. Alice also publishes $S_A$, but signed with Winternitz. All these values are published in the \txname{BobInput} transaction. The script decodes the values $S_A$ and $S_B$, and invalidates the transaction if ($S_B \leq S_A$). Bob’s circuit verifies his proof which checks that (i) $C_B$ score is $S_B$ and (ii) $C_B$ does not contain pegout event PegOutID at position PegOutPos.
        
  \item Secret outputs encode the truth values: Alice’s circuit releases
        \txname{CaTrue}/\txname{CaFalse}, and Bob’s releases \txname{CbTrue}/\txname{CbFalse}.
        Payouts are keyed to these revelations:
        \begin{itemize}
          \item Alice’s deposit is refunded upon \txname{CbFalse} (Bob loses) or after a timelock.
          \item Alice’s deposit is paid to Bob if \txname{CaFalse} (Alice’s proof invalid/loses) or a timelock (Bob’s chain dominates or Alice does not punish Bob).
          \item Bob’s deposit is refunded upon \txname{CaFalse} (Alice loses) or after a timelock.
          \item Bob’s deposit is paid to Alice upon \txname{CbFalse} or after a timelock.
        \end{itemize}
\end{enumerate}

Note that in case neither \txname{CaFalse} nor \txname{CbFalse} are revealed (both circuits evaluate to true), neither party can take each other deposit, but Bob should have been able to. This case we can only handle by slashing a persistent security bond, either stored in Bitcoin or in the side-system.

This \emph{score-carry} construction keeps each circuit close in size to the
non-contestable variant (only one proof is verified per circuit), while still
supporting counter-proofs and an objective fork-choice comparison.

\subsection{Practical Considerations}

We analyze the practical scalability of BATTLE on Bitcoin, with emphasis on the number
of operators the system can support. During setup, each ordered operator pair $(X,Y)$
prepares two garbled circuits (GCs), one for each role direction (``$X$ asserter vs.
$Y$ challenger'' and vice versa), and these artifacts are reused for all peg-ins between
the same pair. Let $S_{\mathrm{gc}}$ denote the ciphertext size of a single GC. The
per-operator storage is dominated by GC material and scales as
\[
\text{storage per operator}\;\approx\;2\,(N-1)\,S_{\mathrm{gc}} \quad \text{(bytes)}.
\]
Under a conservative assumption that a Bitcoin-optimized, 128-bit security DV-SNARK verification garbled circuit requires
$S_{\mathrm{gc}}\in[50,500]\,\mathrm{MB}$ (megabytes), a deployment with $N=1000$
implies
\[
2\,(1000-1)\times[50,500]\,\mathrm{MB} \approx\; [0.1,\,1.0]\,\mathrm{TB}
\]
of GC storage \emph{per operator}\footnote{Although a storage footprint of \(\sim\!1\,\mathrm{TB}\) is moderate, coordinating a setup ceremony with \(N\!\approx\!1000\) participants entails substantial operational risk.}. 

Transaction-DAG metadata and signatures contribute a smaller additive overhead.
 
We now consider the cost of the \emph{transaction DAG}. For each new peg-in, participants
dynamically construct a fresh DAG with $\bigO(N^2)$ pre-signed transactions. These transactions
must be produced, signatures exchanged among parties, aggregated locally by each party,
and the resulting aggregates stored persistently. Consequently, the transaction and
signature counts impose concrete requirements on computation, storage, and network
bandwidth \emph{per peg-in}. Unlike BitVM bridge designs that precompute large DAGs for
future peg-ins, a BATTLE-based bridge amortizes setup by generating the DAG on demand,
thereby avoiding large upfront storage.

Under this dynamic approach, the system can support a large operator set. For example,
with $N=1000$, we estimate per-peg-in overhead (excluding garbled circuits precreated
at setup) of approximately: $\sim$5 minutes of signing (single threaded), $\sim$5 minutes for signature
exchange (peer-to-peer), and $\sim$1\,MB of persistent data per operator until the corresponding peg-out.

Let $U$ denote the number of peg-in UTXOs maintained by the bridge. For $U=1000$ and $N=1000$ parties, the per-party storage required for the pre-signed transaction DAG is approximately $2\,\mathrm{GB}$. At this scale, the storage footprint of pre-signed transactions is of the same order of magnitude as that of the garbled-circuit artifacts.

An additional constraint is the tournament makespan. For a cohort of approximately 1,000 parties, Phase~1 can require up to 60 timelock periods; with a one-day period, this implies a wall-clock duration of about 60 days, which is near the boundary of operational feasibility. In practice, the \AOSB\ implied by input publication is small enough that a concurrency
factor of \(Q=16\) is feasible. At this setting, Phase~1’s wall-clock latency decreases from about \(60\) to \(\sim 36\) timelock periods.

\section{Summary}

We introduce \emph{BATTLE} for Bitcoin, a two-phase tournament that recycles dispute rewards to fund later rounds, keeping the honest asserter’s initial capital constant while resolving $C$ challenges in logarithmic rounds. Phase~1 eliminates competing asserters in a bracket; Phase~2 escalates the number of simultaneous disputes per round according to schedules that trade rounds for liquidity. We give a Bitcoin-native instantiation using \emph{FLEX}-style BitVM garbled-circuit disputes with per-move timelocks, on-demand L1 bonds, escrowed rewards that are immediately reusable, stall handling, and a race-free early-refund gate. We design Phase~1 with enabler chains (cut-by-winner and third-party stall cuts) and a single-winner handoff to Phase~2; we rate-limit openings with a \emph{Tournament Chain} (TC) and analyze the \emph{Open-and-Abandon} attack together with mitigations (operator attribution and slashing, threshold-taproot rate control). We discuss why opener-agnostic on-demand bonds are non-trivial without \texttt{SIGHASH\_NOINPUT}/\texttt{ANYPREVOUT}. For chain evidence, we argue for \emph{contestable} proofs and give two practical constructions (dual-proof input and score-carry) so challengers can defeat non-canonical assertions by supplying higher-work counter-proofs. We also discuss how to fuse FLEX component inputs to amortize on-chain commitments without changing capital bounds, and how to disable losers’ unbonded privileges via per-party secrets (including threshold variants).

Operationally, the pre-signed transaction material scales quadratically in the number of operators, and the overall artifact size is dominated by garbled circuits; dynamic per–peg-in DAG construction spreads costs over time. 

Open problems include tighter mutual-exclusion primitives for tournament admission and sharper, verifiable rate limiters. In sum, BATTLE provides a concrete, Bitcoin-compatible path to DoS-resilient, constant-capital dispute resolution that retains openness while delivering logarithmic settlement time.

\bibliographystyle{IEEEtran}  % or any other style you prefer
\bibliography{references}     % 'references' is the name of your .bib file

\appendix

\end{document}